# Smart Refrigerator using IoT and Android

Abhishek Das[1], Vivek Dhuri[2], Ranjushree Pal[3]

[1,2]U.G. Student, Department of Electronics and Telecommunication, DJSCE, Mumbai, India
[3]Assistant Professor, Department of Electronics and Telecommunication, DJSCE, Mumbai, India
E-mail [1]abhishekdas611 @gmail.com, [2]vivekdhuri867@gmail.com, [3]ranjushree.pal@djsce.ac.in

*Abstract*—**The kitchen is regarded as the central unit of the traditional as well as modern homes.It is where people cook meals and where our families sit together to eat food. The refrigerator is the pivotal of all that, and hence it plays an important part in our regular lives.The idea of this project is to improvise the normal refrigerfator into a smart one by making it to place order for food items and to create an virtual interactive environment between it and the user.**

## I. INTRODUCTION

It is predicted that in the near future almost everything will be connected to the Internet which not only will include PCs, laptops,mobile phones, but also daily utilities such as coffee blenders, refrigerators, washing machines, microwave ovens, juicer, televisions,heaters, air conditioners etc. and have their own IP address . Since when the first smart fridge was introduced in 1990's , researchers have been imagining refrigerators that would change our lifestyles. Smart Refrigerator finds out the available stock of food items such as eggs , milk , jam, sauce etc present in it and sends a sms notification to the user's designated mobile phone number via GSM module if any of these items are finished. It then waits for the user's response (via text sms) on weather to place an order for that particular food item or not. If user replies 'Yes' then it places an order for that item to an designated grocery store again through a SMS along with required time of delivery .If the refrigerator does not gets an command from the user,scanning for that item is paused . It resumes scanning only after the item is replenished back by the user.A threshold level can also be set for quantity of eggs, amount of milk and other items such that when the available stock goes below the threshold, similar action can be taken.

## II. LITERATURE SURVEY

LG started the concept to design an smart refrigerator in 2000 with the Digital DIOS which featured video messaging. [1].Follow it other manufactures such as Samsung ,Panasonic joined the race coming up with concepts of camera integration within the fridge to monitor spoilage and letting us keep an eye on leftover food without opening the door.[2].With the revolution in technology, Samsung has came up with innovative concepts such as integration of touchscreen and mobile applications to order meals,display recipes,Wi-Fi enabled cameras to check supplies and track food expiry date.However these refrigerators are highly expensive and thus are not marketed on a large scale.[3]

## III. BLOCK DIAGRAM

The following diagram gives a brief idea about the model.

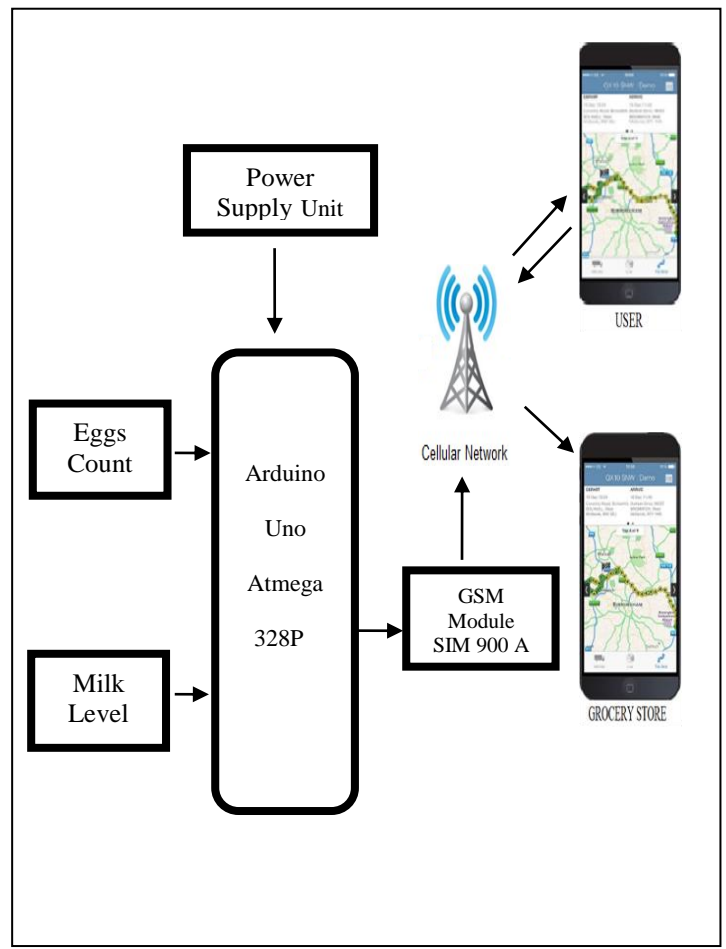

Fig 1. Block Diagram of Proposed Model

## A. GSM SMS Alert System

SIM900A Modem is built with Dual Band GSM/GPRS based SIM900A modem from SIMCOM. It works on frequencies 900/ 1800 MHz. SIM900A can search these two bands automatically.The baud rate is configurable from 1200-115200 through AT command. The GSM Modem is having internal TCP/IP stack which enables you to connect with internet via GPRS. This is a complete GSM/GPRS module in a SMT type and designed with a very powerful single-chip processor integrating AMR926EJ-S core, allowing you to benefit from small dimensions and cost-effective solutions.[4]

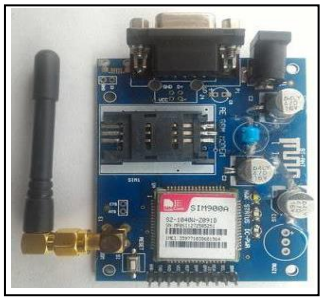

Fig 2. GSM SIM 900A Module

TABLE I. BASIC AT COMMANDS USED

| Command | Description |
| --- | --- |
| AT+CMGD | DELETE SMS MESSAGE |
| AT+CMGF | SELECT SMS MESSAGE FORMAT |
| AT+CMGL | LIST SMS MESSAGES FROM PREFERRED STORE |
| AT+CMGR | READ SMS MESSAGE |
| AT+CMGS | SEND SMS MESSAGE |
| AT+CMGW | WRITE SMS MESSAGE TO MEMORY |
| AT+CMSS | SEND SMS MESSAGE FROM STORAGE |
| AT+CNMI | NEW SMS MESSAGE INDICATIONS |

## B. Arduino Uno Development Board

Arduino Uno is a development board based on a dual-inline-package ATmega328 AVR microcontroller. It has 20 digital i/o pins,6 of it can be used as Pulse Width Modulated(PWM) outputs and 6 can be used as analog inputs.It has a 16 MHz crystal, a USB port, an ICSP header . Programs can be loaded on to it from the Arduino computer program software which is an open source IDE. The Arduino has an vast support community, which makes it a very easy way to get started working with it.

## C. Android Studio 6.0

Android Studio 6.0 is an integrated development environment for Google's Android Operating System . The Android studio is the IDE for app development which provides features and advantages that helps quickly and successfully develop android applications.It is created on JetBrains IntelliJ IDEA software and designed primarily for Android development. It is a replacement for the Eclipse Android Development Tools (ADT) which was primary IDE for Android application development.

## IV. IMPLEMENTATIONN

### A. Eggs Count and Ordering

An Ultrasonic sensor uses sound waves to measure the distance to an object. It sends out a sound wave at a specific frequency and waits for that sound wave to bounce back. By recording the time interval between the sound wave being transmitted and the sound wave received,distance between the sensor and the object is calculated using the basic distance formula as shown in figure 3.An HC-04 Ultrasonic sensor is used in this model .It has a maximum range of upto 400 cm.

In this model, an Ultrasonic sensor is attached to the eggs tray.It continuously emits sound waves which reflect from the eggs placed in front of it and gives a particular distance.It can be programmed with the Arduino such that pre defining different ranges of distance we can calculate the presence of eggs as well as the quantity.Thus we raise an signal if the eggs are depleted or when a minimum threshold quantity is reached.The further operation is carried out by the GSM module and the User.

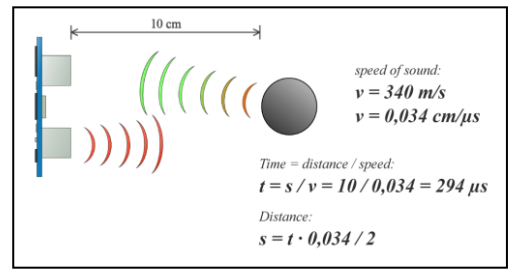

Fig 3. Principle of Ultrasonic Sensor

### B. Milk Measurement and Ordering

The principle of an infrared sensor working as an Object Detection Sensor is used in this proposed part of the project. An infrared sensor consists of an infrared LED and an infrared Photodiode,which is called as Photo Coupler or Opto Coupler. When the transmitter emits radiation, a part of the radiation reflects back from the object if it is present within the range specified to the IR receiver.Based on the intensity of the received signal by the IR receiver, the output value of the sensor is defined.It has a range of 5cm to 30 cm.

The IR unit is attached to the cap of the milk container facing inwards of the container such that it can measure the depth of the inside liquid.We define a level depending on the container size such that if the milk goes below that level i.e the reflected distance becomes large than the threshold set ,an signal is raised and a similar operation of ordering is carried out.This application can be extended to other liquids and solid food items by defining different levels of threshold conditions as per requirement.

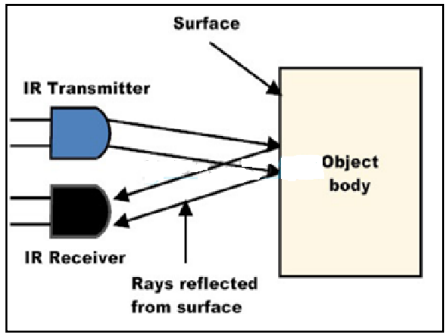

Fig .4. IR Object Detection Principle

*C. Creation of Interactive Interface*

Using the GSM facility, user can also send an inquiry request to the refrigerator about the present stock available in refrigerator.This feature can help in reducing over purchasing of some items while shopping and maintain household budget. Smart Refrigerator can also store information regarding nutritive values of vegetables,related Recipes etc.which the user can avail on demand by sending just an SMS to it. For instance it the user sends a text 'Tomato',then it get a SMS reply containing the nutrition content and google links to certain Recipes.For this we need to create a database in the Arduino for each of the items needed.

*D. Expiry Date Reminder Android App*

Using Android Studio 6.0,an reminder application is made such that when we put a particular food item into refrigerator,by entering it's expiry date in it,we can track the freshness of items and also prevent them from getting spoiled.The reminder intervals can be set according to the user's convenience.. We created three activities and all the activities are connected to each other. The first activity is the main activity, which we see when we open the app. In first activity there is a floating button, it is used to navigate from first activity to second activity. The second activity helps to add details about expiration of food such as its expiry date, time and alarm title. The third activity is used to show all the reminder created by the user and user can also edit the reminders. XML and JAVA are used for building android applications.

*E. Flowchart*

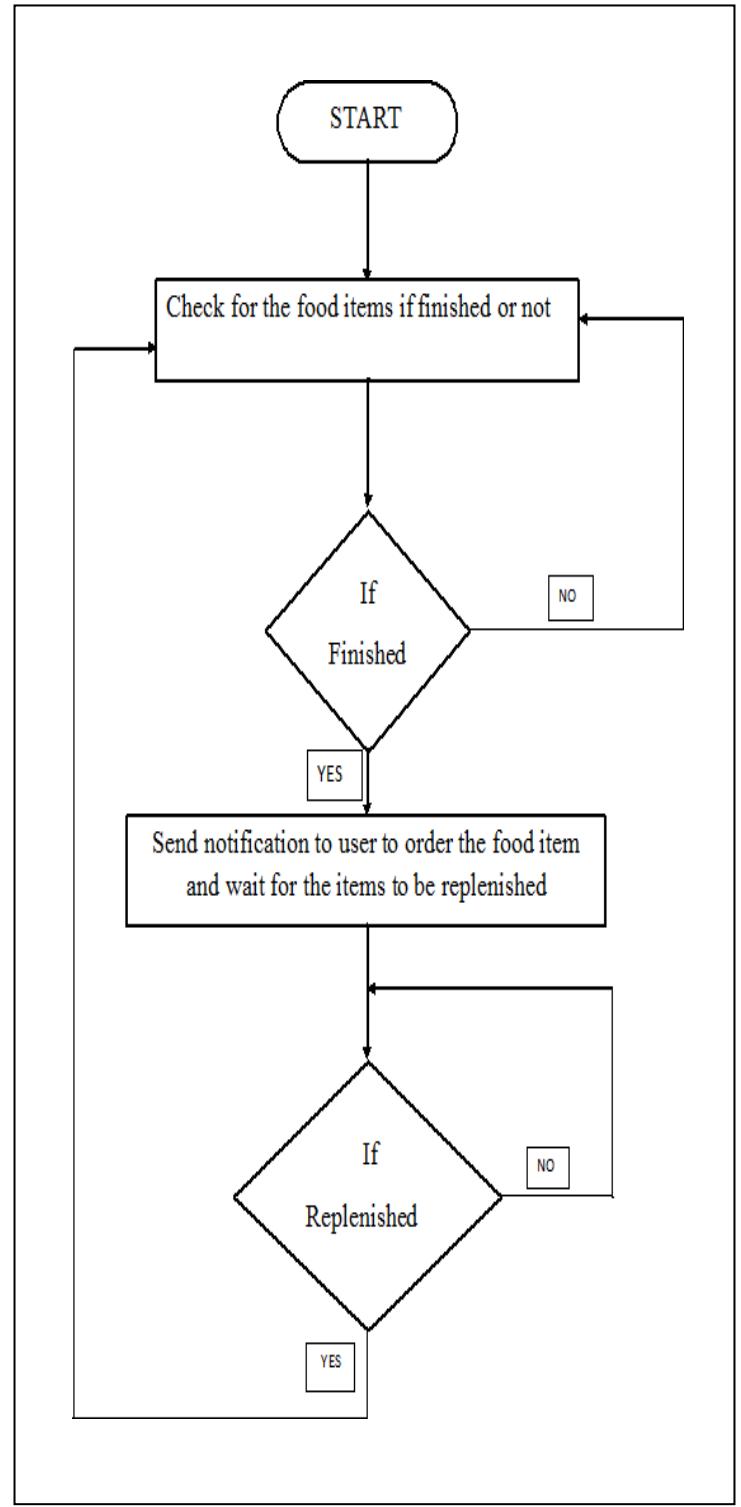

Fig .5. Flowchart

## V. RESULTS

The following are the screenshots of the Food Expiry Date Android Application.

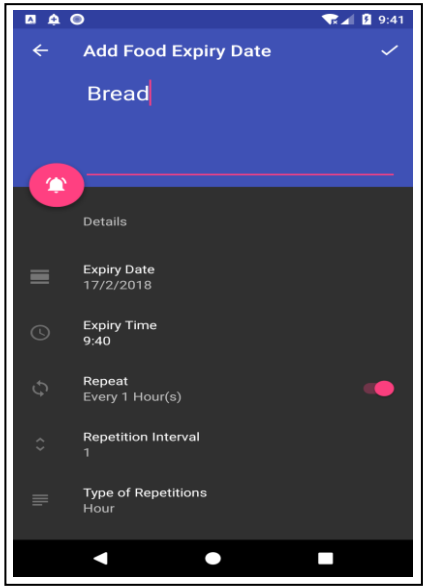

Fig .6. Android Application-1

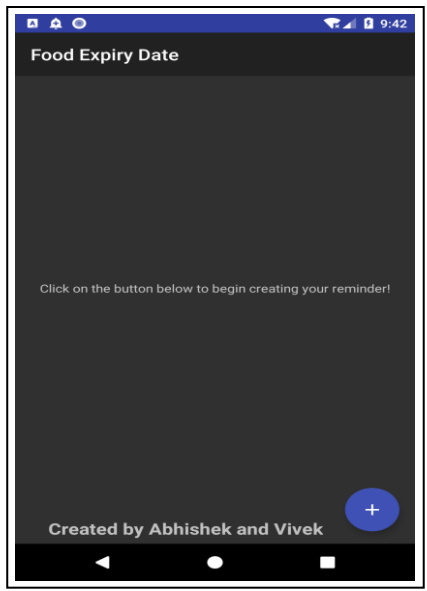

Fig .7. Android Application-2

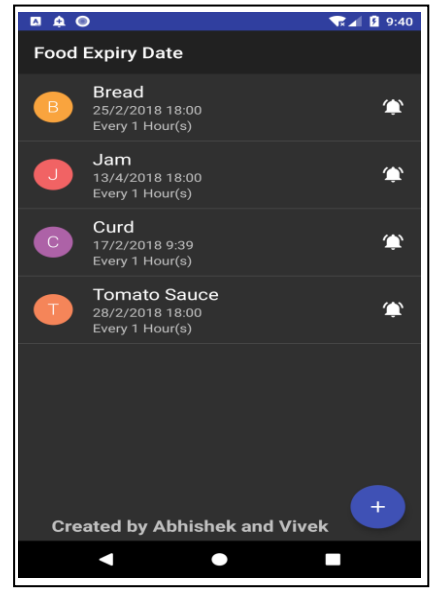

Fig .8. Android Application-3

The following are the screenshots of the messages sent by refrigerator under different situations.

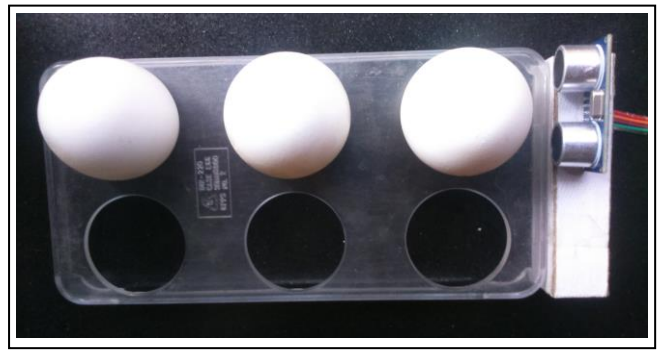

Fig .9. Eggs filled tray

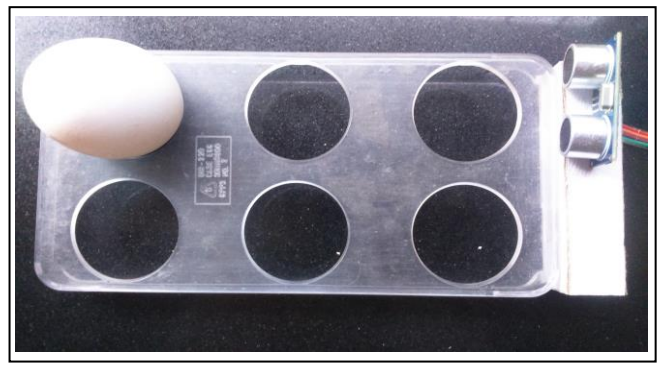

Fig .10. Eggs threshold level reached

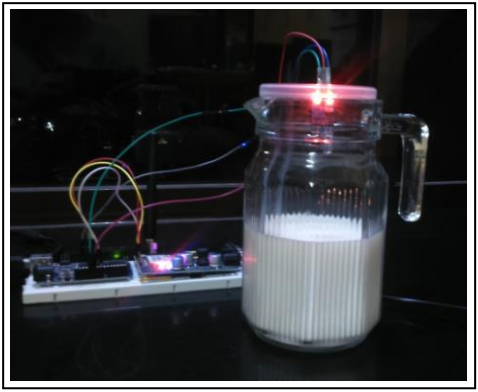

Fig .11. Jar filled with Milk

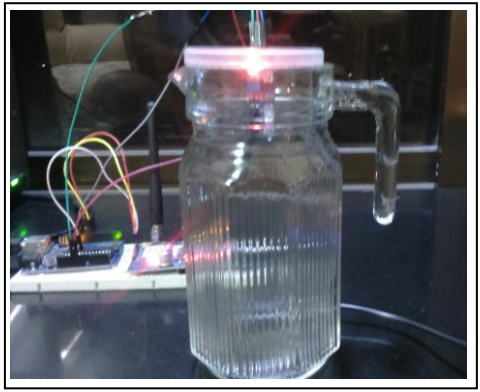

Fig .12. Milk finished condition

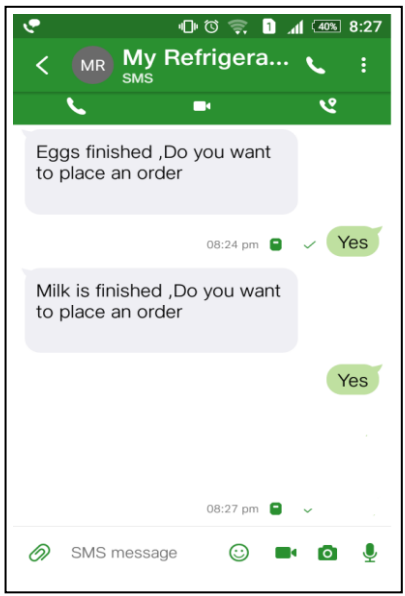

Fig .13.Screenshot of User's Mobile Phone

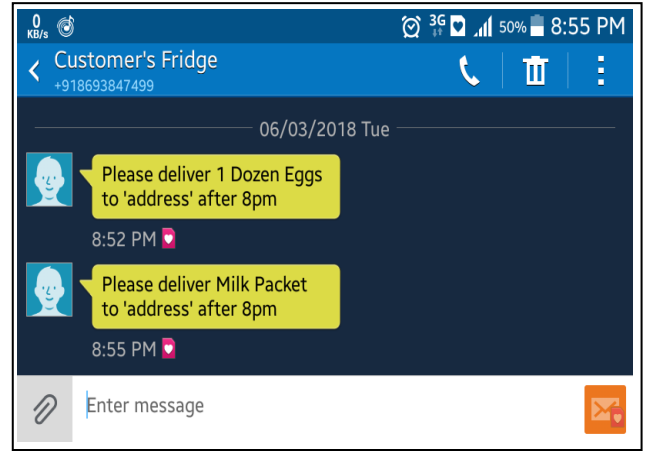

Fig .14. Screenshot of Designated Grocery Store's Mobile Phone

## VI. FUTURE SCOPE AND MARKET POTENTIAL

In future the smart fridge will keep track our food by scanning the its barcodes as we put the items into the fridge.Using voice recognition technology an interactive system can be made,for example, all we have to say is "6 tomatoes" or  "cooked noodles" when placing items into the fridge, which will then record the current date and add an expiration date, so that we never eat spoiled food. With the accelerating growth of the IOT and  Machine Learning techniques in the recent years,these concepts will soon become reality and a part of our lifestyles. Samsung has been planning to launch a regrigerator that has a 21.5-inch LED touchscreen that can display recipes,3 Wi-Fi enabled cameras and an app to check  supplies and track food expiry date.[6]

## ACKNOWLEDGEMENT

We would like to  take this opportunity and thank our head of department Dr. Amit Deshmukh and  faculty guide Ms. Ranjushree Pal  for providing us the competitive platform to implement  such projects.